\documentclass{article}
\usepackage[english]{babel}
\usepackage[utf8]{inputenc}
\usepackage{bm}
\usepackage{amsmath,amsthm,amsfonts}
\usepackage{graphicx}
\usepackage{color}
\usepackage{multirow}
\usepackage{subcaption}
\usepackage{fancyhdr}  
\usepackage{geometry}  

\graphicspath{ {./figs/} }

\date{}

\newcommand{\mytitle}{Physical Similarity of Fluid Flow in Bimodal Porous Media: \\ Part 1 - Basic Model and Solution Characteristics}

\title{\mytitle \thanks{Part of this work was conducted by the authors at Texas A\&M University's Qatar branch campus.}}

\author{
Yuhe Wang \thanks{National \& Local Joint Engineering Laboratory for Big Data Analysis and Computing Technology, Beijing 100190, China.}
\thanks{Institute for Scientific Computation, Texas A\&M University, College Station, Texas 77843, USA. Email: {\tt yuhe.wang@tamu.edu}.}
\and
Yating Wang \thanks{School of Mathematics and Statistics, Xi’an Jiaotong University, Xi’an, Shaanxi 710049, China.}
}

\date{\normalsize \today}

\begin{document}

\maketitle

\thispagestyle{empty}

\pagestyle{fancy}  

\begin{abstract}
Fluid flow through bimodal porous media, characterized by a distinct separation in pore size distribution, is critical in various scientific and engineering applications, including groundwater management, oil and gas production, and carbon sequestration. This note delves into the physical similarity of fluid flow within such media, bridging the gap between microscale phenomena and macroscale observations. We present a representative mathematical model that conceptualizes bimodal porous media as a double-continuum system, distinguishing between macroporous and microporous regions. The model captures the complex interactions between these regions, particularly focusing on the challenges of modeling fluid flow when there is significant disparity in pore sizes. By employing a heuristic approach grounded in pore-scale tomography, we derive governing equations that describe fluid flow and analyze the solution characteristics. The results reveal unique features of the fluid flow in bimodal systems, such as the occurrence of boundary discontinuities and the delayed transient response, which are not observed in conventional porous media. This work provides ground for further studies in bimodal porous media, offering insights that could enhance predictive modeling and optimization in various applications concerning porous media with similar bimodal pore size distributions.  \\ \\
\noindent\textbf{Keywords:} Bimodal porous media, Fluid flow through porous media, Microscale, Macroscale, Physical similarity, Double-continuum model
\end{abstract}

\section{Background}

\begin{figure}[h!]
\centering
\includegraphics[width=0.5 \textwidth]{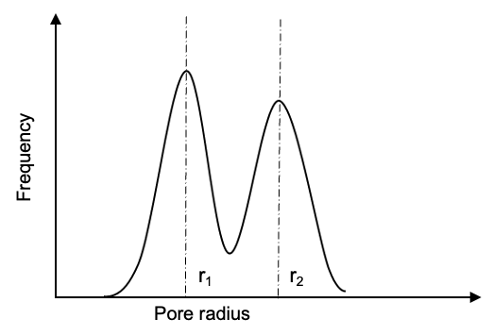}
\caption{An illustration of a typical bimodal pore size distribution}
\label{fig:biomdal}
\end{figure}

Sharp bimodal pore size distributions are frequently observed in both naturally occurring and artificially fabricated porous media, including natural rocks \cite{song2000determining, wang2020generalized,mi2017enhanced, yan2018enhanced}, zeolites \cite{na2011directing}, photocatalytic TiO2 \cite{lee2008synthesis}, and porous carbons \cite{he2011high}. The fluid flow processes within these bimodal porous media are of considerable scientific and engineering importance. A typical sharp bimodal pore size distribution is characterized by two distinctly separated peaks, as illustrated in Figure \ref{fig:biomdal}. The magnitude difference between the two peak pore radii can be the order of two or more, as is often the case for carbonate rocks \cite{song2000determining, wang2020generalized, moshier1989microporosity}. Bimodal carbonates, commonly referred to as microporous carbonates, consist of macropores with radii in the range of hundreds of microns and micropores with radii of a few microns. Understanding the fluid flow through microporous carbonates is fundamental to several critical applications, including groundwater management \cite{budd1989micro, chen2004groundwater}, oil and gas production from carbonate reservoirs \cite{pak2015droplet, zhang2018multiscale, zhang2019novel}, and carbon capture and subsurface storage \cite{lisbona2010integration, falkowski2000global}. 

The bimodal porous structure leads to bimodal permeabilities in the microporous system \cite{nelson1994permeability}. If we decompose the porous bulk into two portions – microporous and microporous portion, the permeability of the microporous portion is often in a few hundreds of milli-Darcy, while the microporous portion often exhibits permeability in just a few milli-Darcy \cite{anselmetti1998quantitative}. This results in a distinct difference in the flow capacity or conductivity of the two systems. In a typical porous medium, the pore space formed by solid grains facilitates flow conductivity. However, in microporous carbonates, the grains are filled with micropores rather than being absolute solid, a feature linked to their geological origins and observed by computed tomography imaging of rock samples \cite{remeysen2008application, sok2010pore, santini2017carbonate, norbisrath2015electrical}.

To understand the fluid flow phenomena in a bimodal porous medium like the microporous carbonates, it is evident that one cannot simply ignore the large portion of micropores. The significant pore size discrepancy introduces substantial computational challenges. At a scale where detailed macropores can be resolved, one can relatively easily solve the Stokes flow through the microporous portion if the contribution of micropores in grains is ignored. However, to account for the micropores, it would be necessary to zoom in on every grain to resolve these micropores and solve a coupled system, which could present enormous difficulties at the boundary between macropores and micropores. Even if one were able to overcome the challenges of modeling the detailed fluid flow in a zoom-in region of microporous carbonates, it would still be impossible to capture the full spectrum of pore-scale configurations of a macroscale system, even with the most advanced tomography technology. The level of detail in imaging is inversely proportional to the sampling size, resulting in extremely sparse and incomplete information on a macroscale system. 

Microscale imaging, dynamic imaging, and modeling provide a viable approach for obtaining direct information at microscale or pore scale. However, it is essential to understand how to scale these microscale results to the macroscale. Detailed microscale modeling of fluid flow may be of little value if scaled learning or mean characteristics of the media and flow cannot be extracted. Therefore, the concept of physical similarity or scaled learning should not be overlooked in the pursuit of microscale modeling. Indeed, the theory of fluid flow has evolved and should continue to evolve toward obtaining the mean characteristics of the media and flow, and formulating the basic laws based on these mean characteristics. Whether in regular or microporous systems, this approach is important toward understanding and modeling fluid flow in porous media. While recent advances in data-driven methods using various machine learning techniques have shown great promise in aiding analysis \cite{samnioti2023applications1, samnioti2023applications2, du2020connectivity, menke2021upscaling}, it is vital that physics remains central when applying such approaches to studies of this nature \cite{yan2022physics, du2023novel}. 

In pursuit of these mean characteristics and scaling laws for physical similarity, we work on a simple mathematical model in this part to describe the fluid flow process through bimodal porous media. We discuss the basic laws and introduce the mean characteristics derived from the mathematical model. This work in modeling fluid flow through bimodal porous media is to assist our understanding of fluid mechanics in such media and have many implications in various scientific and engineering applications as discussed above. 

In this part, we present a mathematical model based on the assumption of a specific microporous structure. We outline the basic concepts of the model. We demonstrate the journey to the self-similar solution. We present and discuss the mathematical equations and the solution characteristics. 

\section{Basic model}

We start by introducing a heuristic model for fluid flow through bimodal porous media. This model is based on pore-scale tomography images. Figure 2 shows a conceptual illustration of bimodal porous media at the pore scale. We assume the porous space is divided into two portions: one representing macropores and another consisting of inclusions with micropores. These microporous inclusions simplify the concept of microporous grains. In microporous carbonates, macropores have a pore size of a few hundred microns with permeability around a few hundred milli-Darcy. Micropores, on the other hand, have a pore size of a few microns with permeability around a few milli-Darcy.  

\begin{figure}[h!]
\centering
\includegraphics[width=0.5 \textwidth]{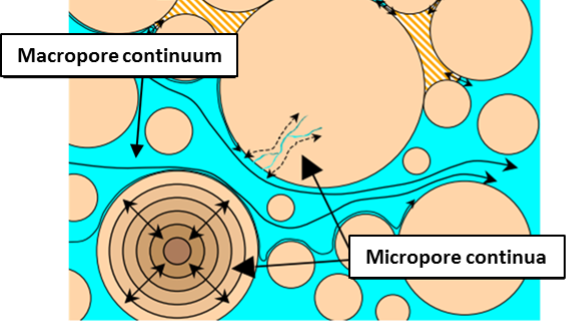}
\caption{An illustration of a typical microporous example, adopted from \cite{lichtnerrole}}
\label{fig:microporous}
\end{figure}

In this basic model, we refer to the microporous continuum as the primary medium and the microporous inclusions as the secondary medium (Figure  \ref{fig:microporous}). Inspired by the pioneering work on fluid flow in fissured rocks \cite{barenblatt1960basic}, we develop a double-continuum model for porous media with microporous inclusions. At this stage, the model operates at the pore scale or microscale. Given the low Reynolds number, it is reasonable to assume that flow in the primary medium follows Stokes flow.  

The basic model introduces two fluid pressures at the same spatial point: one for the primary medium and one for the secondary medium. We also account for the transfer between these two media. While the basic model has a heuristic foundation, the subsequent analysis is rigorous. This heuristic approach includes several key points essential for understanding the rigorous analysis and solutions that follow.

\section{Double-continuum model equations}
Based on mass conservation, we can write two equations as follows: one for the primary medium and another for the secondary medium.

\begin{equation}
\label{dm1}
    \frac{\partial \phi_1 \rho}{\partial t} + \text{div}(\rho u) = \rho \sigma k_s \frac{\mu (p_2 - p_1)}{\mu}
\end{equation}
\begin{equation}
\label{dm2}
    \frac{\partial \phi_2 \rho}{\partial t} = \rho \sigma k_s \frac{\mu (p_1 - p_2)}{\mu}
\end{equation}

where $\rho$ is the fluid density, $\phi_1$ and $\phi_2$ are the porosities of the primary and secondary medium, $u$ is the fluid velocity, $\sigma$ is a factor characterizing the geometry of the bimodal porous media, $k_s$ is a permeability relating to the transfer capability between the two media, $\mu$ is the fluid viscosity, $p_1$ and $p_2$ are the fluid pressures of the primary and secondary medium. The flow in the primary medium at pore scale is assumed to follow Stokes, such that we have 

\begin{equation}
\label{stokes}
    -2\mu \nabla \cdot \mathbf{D}u + \nabla p_1 = 0
\end{equation}

Due to the low permeability of the secondary medium and its representation as microporous inclusions, we disregard the flow within the secondary medium itself. It only contributes to the flow in the primary medium in a slow and steady-state manner. This basic mathematic model has several interesting aspects. Notably, the assumption implies that a spatial point in the secondary medium connects only with the corresponding point in the primary medium, with no connections defined within the secondary medium. 

Rewriting in terms of fluid pressure $p_1$ and $p_2$, we have

\begin{equation}
\label{dm11}
    \phi_1 c_1 \frac{\partial p_1}{\partial t} + \nabla \cdot u = \frac{\sigma k_s}{\mu} (p_2 - p_1)
\end{equation}

\begin{equation}
\label{dm12}
    \phi_2 c_2 \frac{\partial p_2}{\partial t} = \frac{\sigma k_s}{\mu} (p_1 - p_2)
\end{equation}

where \( c_1 \) and \( c_2 \) are the compressibilities of the primary and secondary medium.

After combining Equations \eqref{dm11} and \eqref{dm12} to eliminate \( p_2 \), we have:

\begin{equation}
\label{dm3}
    \frac{\mu}{\sigma k_s} {\phi_1 c_1} {\phi_2 c_2} {\frac{\partial^2 p_1}{\partial t^2}} + (\phi_1 c_1 + \phi_2 c_2) \frac{\partial p_1}{\partial t} + \nabla \cdot (u + \frac{\mu}{\sigma k_s} \phi_2 c_2 \frac{\partial u}{\partial t}) = 0
\end{equation}

At first glance, one can see the similarity between Equation \eqref{dm3} and the governing equation for ordinary porous media flow. We can interpret Equation \eqref{dm3} as an apparent mass balance equation for flow in bimodal porous media.

Additionally, we can define an apparent velocity as:

\begin{equation}
    u_a = u + {\frac{\mu}{\sigma k_s}} {\phi_2 c_2} {\frac{\partial u}{\partial t}}
\end{equation}

It should be noted that $\sigma$ carries the geometrical characteristics of the bimodal media and is inversely proportional to $l^2$, where $l$ is the characteristic length or size of the secondary medium. $\sigma k_s$ characterizes the intensity of fluid transfer between the secondary and primary media, depending on the permeability of the microporous inclusion and the extent of inclusion in the bimodal media. One can think of $\sigma$ as a measure of the specific surface area of the contact between pore and microporous inclusion. As $\sigma$ approaches infinity, it corresponds to a reduction in the dimension of the secondary medium and an increase in the portion of the primary media. Equation \eqref{dm3} then tends to coincide with the ordinary equation for fluid flow in porous media, and the apparent velocity approaches the true velocity. 

\section{Properties of the basic model}

The double-continuum model exhibits peculiar features due to the nature of the system described by Equation \eqref{dm11} and \eqref{dm12}. These uncommon features relate to a medium that contains inclusions with relatively negligible permeability but a large storage volume. To understand this, we first analyze its boundary value properties. In the context of fluid flow, a homogeneous initial condition is of primary interest. Our analysis focuses on two key types of boundary conditions: Dirichlet and Neumann. We further narrow the discussion to a case of particular engineering interest, where the initial and boundary conditions are unrelated. One can observe a peculiar behavior in the existence of boundary discontinuities in pressure $p_1$ and the first-order spatial derivative of $p_1$, which cannot be eliminated instantaneously. This is a distinct characteristic of the double-continuum model compared to the ordinary porous media flow model.   

\section{Discontinuities of boundary value}

Considering a surface at \(x=0\), let’s examine the one-dimensional case with a sufficiently small vicinity \(S = -h \leq x \leq h\), \(0 \leq t \leq T\) on both sides of this surface. The variable \(x\) follows the normal direction to this surface. \(h\) is a small quantity and \(T\) represents an arbitrary time. We simplify our analysis by applying the Hagen-Poiseuille equation \(u = -\frac{R^2}{8\mu} \frac{\partial p_1}{\partial x}\) to describe the Newtonian laminar flow crossing this surface \cite{wu2017wettability, song2018engineering}. \(R\) is the radius of the 1D flow in the dimension perpendicular to \(x\). With these considerations, we can write Equation \eqref{dm3} as:

\begin{equation}
\label{lp1}
    L_{p_1} := \alpha_1 \frac{\partial^2 p_1}{\partial t^2} + \alpha_2 \frac{\partial p_1}{\partial t} - \beta_1 \frac{\partial^2 p_1}{\partial x^2} - \beta_2 \frac{\partial^3 p_1}{\partial x^2 \partial t} = 0
\end{equation}

where:

\[
    \alpha_1 = \frac{\mu} {\sigma k_s} {\phi_1 c_1 \phi_2 c_2}, \quad \alpha_2 = \phi_1 c_1 + \phi_2 c_2, \quad \beta_1 = \frac{R^2}{8 \mu}, \quad \beta_2 = \frac{\phi_2 c_2 R^2}{8 \sigma k_s}.
\]

We further assume that \(p_1\) is continuous to satisfy Equation (8). By applying integration by parts, we obtain:

\begin{equation}
\label{p1}
   \left|p_1\right| = \left|p_1\right|_{t=0} e^{-\frac{\beta_1 t}{\beta_2}}
\end{equation}

\begin{equation}
\label{dp1}
    \left|\frac{\partial p_1}{\partial x}\right| = \left|\frac{\partial p_1}{\partial x}\right|_{t=0} e^{-\frac{\beta_1 t}{\beta_2}}
\end{equation}

where $\left| \cdot \right|$ denotes the difference between the values on both sides of the discontinuity surface. The derivations of Equations \eqref{p1} and \eqref{dp1} are omitted to keep the content concise and they are available upon request.

In its physical sense, the jump in pressure as indicated by Equation \eqref{p1} and the jump in the normal derivative of pressure as shown in Equation \eqref{dp1}, caused by the discontinuities or inconsistent initial conditions, decay according to the law \(e^{-\frac{\beta_1 t}{\beta_2}}\), rather than being eliminated instantaneously as in an ordinary porous medium. One should also note that the normal derivative of pressure is proportional to fluid velocity \(u\). We denote this property as a characteristic of the mathematical description of fluid flow in bimodal porous media. The jump tends to disappear as time approaches infinity for bimodal porous media. For a fixed time, one can observe that this jump also tends to vanish as \(\sigma\) goes to infinity. Indeed, in the limiting case where \(\sigma\) is infinite, bimodal porous media behave like ordinary porous media. This is an important property of this boundary value problem. Although there is a jump in velocity at the boundary, the total flow or apparent velocity remains continuous. This finding is critical when seeking solutions under constant velocity conditions. This property suggests that one should use the apparent velocity to impose the boundary velocity condition.

\section{Solution Characteristics}

We then seek to determine the solution characteristics of the model equation Equation \eqref{lp1} under the following conditions:

\begin{align} \label{cond1}
    p_1(x, 0) &= p_0 \quad 0 \leq x < \infty, \nonumber \\
    p_1(0, t) &= p_b \nonumber \\
    p_1(\infty, t) &= p_0
\end{align}

Equation \eqref{cond1} says that at the initial instant there is a fluid pressure change of \((p_0 - p_b\)) at the boundary. Then, the fluid pressure immediately to the right of the boundary is equal to:

\[
p(0^+, t) = p_b + (p_0 - p_b)e^{-\frac{\beta_1 t}{\beta_2}}.
\]

To find the pressure distribution at any instant of time \(t\), we let \(h(x, t) = \frac{p_0 - p_1(x,t)}{p_0 - p_b}\), where \(h(x,t)\) satisfies:

\begin{align} \label{cond2}
    \alpha_1 \frac{\partial^2 h}{\partial t^2} + \alpha_2 \frac{\partial h}{\partial t} - \beta_1 \frac{\partial^2 h}{\partial x^2} - \beta_2 \frac{\partial^3 h}{\partial x^2 \partial t} &= 0, \quad 0 \leq x < \infty, \, t > 0, \nonumber \\
    h(x, 0) &= 0, \quad 0 \leq x < \infty, \nonumber \\
    h(0, t) &= 1 + e^{-\frac{\beta_1 t} {\beta_2}}, \quad t > 0, \nonumber \\
    h(\infty, 0) &= 0
\end{align}

Applying Laplace transformation and residue theorem (derivation available upon request), we obtain:

\begin{equation} 
\label{eq13}
h(x, t) = 1 - \frac{1}{\pi} \int_0^1 e^{-\frac{\beta_1 t}{\beta_2}} \sin(\sqrt{\frac{\lambda(1 - c\lambda)}{1 - \lambda}}\sqrt{\frac{\alpha_2}{\beta_2}}) \frac{d\lambda}{\lambda(1-\lambda)}
\end{equation}

where \(c = \frac{\alpha_1 \beta_1}{\alpha_2 \beta_2}\).

Let \(z = \frac{\beta_1 t}{\beta_2}\) and \(\xi = \sqrt{\frac{\alpha_2}{\beta_1}}\frac{x}{\sqrt{t}}\), we arrive at the self-similar solution in terms of:

\begin{equation} 
\label{eq14}
    h(x, t) \Rightarrow h(\xi) = 1 - \frac{1}{\pi} \int_0^1 e^{-z\lambda} \sin(\sqrt{\frac{\lambda(1 - c\lambda)}{1 - \lambda}}\xi\sqrt{z}) \frac{d\lambda}{\lambda(1 - \lambda)}
\end{equation}

We can further change the integration limit from \(0,1\) to \(0,\infty\) as shown in:

\begin{equation} 
\label{eq15}
    h(x, t) = 1 - \frac{1}{\pi} \int_0^\infty e^{-\frac{\beta_1 v^2}{\alpha_2 + \beta_2v^2}t} \sin(vx\sqrt{\frac{\alpha_2 + \beta_2(1 - c)v^2 }{\alpha_2 + \beta_2v^2}})\frac{2dv}{v}
\end{equation}

Then, Equation \eqref{eq15} becomes:

\begin{equation} 
\label{eq16}
    h(\xi) = 1 - \frac{1}{\pi} \int_0^\infty e^{-\frac{v^2}{4\xi^2 + \frac{\beta_2}{\beta_1}v}t} \sin(v\sqrt{\frac{4 \xi^2 \beta_1 t + \beta_2(1 - c )v^2}{4 \xi^2 \beta_1 t + \beta_2 v^2}})\frac{2dv}{v}
\end{equation}

Since \(\beta_2 = \frac{\phi_2 c_2 R^2}{8 \sigma k_s} \propto \frac{1}{\sigma}\), we can see \(\beta_2\) tends to 0 as \(\sigma\) approaches infinity, which clearly corresponds to a dimension reduction of the secondary medium and an increase of the portion of the primary medium. When \(\beta_2 \rightarrow 0\), Equation \eqref{eq13} becomes:

\begin{equation} 
\label{eq17}
    h(x, t) = 1 - \frac{2}{\pi} \int_0^\infty e^{-\frac{\beta_1}{\alpha_2} v^2 t} \sin(vx) \frac{dv}{v}
\end{equation}

If we then let \(u = vx\) and \(\xi = \frac{x}{2\sqrt{\frac{\beta_1}{\alpha_2}t}}\), Equation \eqref{eq17} can be written as:

\begin{equation} 
\label{eq18}
    h(x, t) = 1 - \frac{2}{\pi} \int_0^\infty e^{-\frac{u^2}{4\xi^2}} \frac{\sin(u)}{u} du
\end{equation}

Note that in Equation \eqref{eq15} we apply the same self-similar variable \(\xi\) up to a scaling scalar. The well-known self-similar solution of the classical porous media flow equation is obtained from Equation \eqref{eq16}. In the error function form we have:

\[
h(x, t) = 1 - \frac{2}{\pi} \, \text{erf}(\xi), \quad \text{or} \quad p_1(x,t) = p_b + \frac{2}{\pi}(p_0 - p_b) \, \text{erf}(\xi)
\]

\section{Concluding remarks}

In Part 1, we provide a simple mathematical model for fluid flow in bimodal porous media. We demonstrate the process of reaching a self-similar solution at microscale. We discuss the discontinuities of the boundary value and the solution characteristics, which could provide valuable insights. When dealing fluid flow processes in bimodal porous media, the ordinary equations of porous media flow can be applied only if the characteristic times of the process under consideration are long compared to the delay time. However, if it is not the case, the basic dual-continuum model presented in this note may be necessary. In many instances, it is crucial to consider the secondary pores when investigating such processes. In the next part of this study, we provide the homogenization of the basic model to Darcy scale in the classic pressure transient analysis context of petroleum reservoir studies. We also discuss previously unrevealed flow behaviors and the related engineering implications. 

Some intermediate steps in the derivation of the solutions have been omitted. For those details, please contact the corresponding author.    

\bibliographystyle{unsrt}
\bibliography{bimodal-1}

\end{document}